\documentclass[sigconf]{acmart}

\usepackage{graphicx}
\usepackage{amsmath}
  
\usepackage{amssymb}
\newcommand{\R}{\mathbb{R}}
\usepackage{multirow}
\usepackage{tabto}
\usepackage{booktabs} 
\usepackage{algorithmic}
\usepackage{algorithm}
\usepackage{graphicx}
\usepackage{subcaption}

\usepackage{float}
\AtBeginDocument{%
  \providecommand\BibTeX{{%
    \normalfont B\kern-0.5em{\scshape i\kern-0.25em b}\kern-0.8em\TeX}}}

\renewcommand\footnotetextcopyrightpermission[1]{} 

\begin{document}
\setlength{\textfloatsep}{5pt}
\setlength{\intextsep}{1pt}
\title{Hawkes Process Classification through Discriminative Modeling of Text}

\author{Rohan Tondulkar}
\email{cs17mtech11028@iith.ac.in}
\affiliation{%
  \institution{IIT Hyderabad}
}

\author{Manisha Dubey}
\email{cs17resch11003@iith.ac.in}
\affiliation{%
  \institution{IIT Hyderabad}
}

\author{P.K. Srijith}
\email{srijith@cse.iith.ac.in}
\affiliation{%
  \institution{IIT Hyderabad}
}

\author{Michal Lukasik}
\email{mlukasik@google.com}
\affiliation{%
  \institution{Google Inc.}}


\begin{abstract}
Social media has provided a platform for users to gather and share information and stay updated with the news. Such networks also provide a platform to users where they can engage in conversations. However,  such micro-blogging platforms like Twitter restricts the length of text. Due to paucity of sufficient word occurrences in such posts, classification of this information is a challenging task using standard tools of natural language processing (NLP). Moreover, high complexity and dynamics of the posts in social media makes text classification a challenging problem.  However, considering additional cues in the form of past labels and times associated with the post can be potentially helpful for performing text classification in a better way. To address this problem, we propose models based on the  Hawkes process  (HP)  which  can  naturally  incorporate  the temporal  features  and  past  labels  along  with  textual  features for improving short text  classification. In particular, we propose a discriminative approach to model text in HP where the text features parameterize the base intensity and/or the triggering kernel. Another major contribution is to consider kernel to be a function of both time and text, and further use a neural network to model the kernel. This enables modeling and effectively learning the text along with the historical influences for tweet classification. We demonstrate the advantages of the proposed techniques on standard benchmarks for rumour stance classification.
\end{abstract}




\maketitle

\section{Introduction}
Social media platforms like Twitter, Facebook, WhatsApp etc. provide platform for common users to share information and content. However, content shared on these websites are not verified and misinformation or rumours can spread quickly through social media. Social media has become the starting point for many rumours and fake news.  In India, recently a rumour about a child-lifting gang was the cause of death of 29 people\footnote{https://theprint.in/opinion/a-single-whatsapp-rumour-has-killed-29-people-in-india-and-nobody-cares/77634/}. False rumours about death of famous celebrities like Justin Bieber, Charlie Sheen, Taylor Swift, etc have been common in the past. During the English riots in 2011, false rumours about various incidents were spread via Twitter and Facebook. We have encountered incidents where rumours and fake news are used in social media to influence the outcome of elections. 
Rapid spread of rumours through social media can create chaos and cause lot of damage to life and property. Thus, it is very important to keep the spread of rumours in check. Analyzing various aspects of a rumour can help in reducing the damage. If correct information is provided to authority related to rumour, it can help them in taking corrective measures sooner.

However, social media networks provide a platform for common users to share information, generally in the form of short snippets of text, with a prominent example of Twitter. The effective involvement of such information pieces can be useful in addressing various real world problems. For instance, it can help in understanding the stance or opinion of people towards a product, or even prevent the spread of rumours through rumour stance classification ~\cite{zubiaga2018discourse}.
However, tweets involve frequent use of informal grammar as
well as irregular vocabulary e.g. abbreviations, typographical errors and hashtags. The posts exchanged via Twitter are referred to as micro-blogs
because there is a 140 character limit imposed by Twitter for every tweet. Since, these texts are short in nature, therefore do not provide sufficient word occurrences.
This often acts as limitation for classification of social media posts.
However, considering additional cues in the form of past labels and times associated with the social media posts can help in more effective classification thereof.

Motivated with this concern, our work targets the problem of rumour stance classification. In rumour stance classification task, we classify the posts following a would-be rumour post as supporting, denying, questioning or commenting about the rumour. During the spread of rumours via social networking platforms like Twitter, a previous tweet can influence a response in the form of another tweet.  This process continues and various tweets, replies and retweet events are formed in a short time until a cool off period is reached. Such characteristics like cluster of events and self-excitation can be modelled using Hawkes process~\cite{hawkes1974cluster} (HP) model. The stance associated with a post depends on the labels associated with the past tweets and the time of those posts.  This can be naturally and easily considered using a HP model and  makes it a suitable candidate  to solve the problem of stance classification in social media \cite{rizoiu2017tutorial}.

We propose a discriminative modeling of text where textual features are a part of the intensity function of HP. This allows HP to consider the impact of  text as well as time. For text classification problems, discriminative methods were found to perform better \cite{jebara1999maximum}. 
The kernel in a HP model represents the influence of historical events on current event. Our work provides new direction on how various kernels can be used to consider historical impact of text along with time. We show ways in which neural networks can be used instead of standard kernels to learn complex non-linear relationships of influence from historical events. The proposed Neural Kernel Hawkes process provides dual benefit i.e. to use the power of neural networks for learning non-linear relationships and  still maintain the explainability provided by HP. 
\paragraph{\textbf{Contributions}} Our contributions can be summarized as follows:
\begin{itemize}
\item We propose discriminative modeling of text using Hawkes process.
\item We propose use of various text-based kernels to understand historical influence of text in determining stance. 
\item We propose the use of neural networks as a kernel in HP intensity function to learn complex non-linear relationships between events.
\item We show improvement in performance using the proposed models in rumour stance classification problem. 
\end{itemize}

Although we have performed this task of text classification for rumour stance classification, it is a general method which can be used for various other applications like rating prediction etc.



\section{Related Work}
\subsection{Rumour Stance Classification}
Stance classification problems in social media were tried to solve using tree based structures like Linear-Chain conditional random field (CRF) and Tree CRF~\cite{zubiaga2016stance}. \cite{aker2017simple}  used various machine learning classifiers using  problem-specific features and provided insight on whether it was necessary to use more complex models or extract better features.  A long short term memory (LSTM) based sequential approach was used in \cite{kochkina2017turing} that modelled the conversational structure of tweets. With new approaches being developed in deep learning, similar were applied to the problem of rumour stance classification. Attention models using convolution neural networks and LSTMs were used for multi-class classification with sequence of threads of tweets as input in \cite{veyseh2017temporal}. They also used follower-followee relationship between users as an added feature. Four different sequential classifiers using LSTM were used on local and contextual features \cite{zubiaga2018discourse} which showed the higher performance of LSTM. In \cite{kochkina2018all}, a multi-task learning model is proposed which uses a shared LSTM to solve, rumour detection, stance classification and veractiy prediction, all together to also learn common characteristics. \cite{santosh2019can} proposes a Siamese adaptation of LSTMs with an attention mechanism for stance classification problem. The usefulness of LSTM and other sequence models for stance classification of short texts shows that considering past labels will be useful to solve stance classification effectively. This is especially useful in classifying short text arising in social media.

There exist few works in literature where the Hawkes process is used for natural language modeling like topic modeling~\cite{he2015Hawkestopic}, clustering document streams~\cite{du2015dirichlet}, discovering topical interactions~\cite{bedathur2018discovering} and narrative reconstruction~\cite{seonwoo-etal-2018-hierarchical}. 
Although there have been works in the intersection of Hawkes process and topic modeling for various applications like detecting fake retweets~\cite{Dutta2020HawkesEye} and modeling of COVID-19 Twitter narratives~\cite{sha2020dynamic}, incorporating text in the framework of Hawkes process has not been extensively studied. 

A closely related approach for stance classification is~\cite{lukasik2016hawkes} where authors have used multivariate Hawkes Process (MHP) to solve this problem using a generative approach of modeling text. They used a Hawkes process model which consider both  labels and time of past posts along with text to perform stance classification. The text features are  considered through an additional likelihood. A multinomial distribution is used to model text, where the likelihood of generating text is given by 
$
    p(X_n | y_n) =  \prod_{v=1}^V \beta_{y_n v}^{X_{nv}}
$. Here, $V$ is the number of words in vocabulary and $\beta$ is the matrix of size $|Y| \times V$ providing the word distribution for each class. However, this generative model is restrictive  and prevents consideration of text in determining the influence from the past events. For instance, posts with similar textual content will have higher influence in determining the stance of the current post. 

In view of this, we propose HP models which consider text in a discriminative manner. We include various ways to incorporate text within the intensity function of Hawkes process. Therefore, our proposed approach perform time sensitive sequence classification of a tweet by considering label, time and text associated with the previous tweets through  intensity function. This leads to more powerful HP models which can perform stance classification considering the influence not only from past labels and time, but also from text. Moreover, we also provide more generic way to model the influence using a neural kernel.   

\section{Problem Statement}
We consider tweets associated with $D$ topics (or statements or claims) of interest   for stance classification. 
Each tweet is represented as a tuple $d_j = (t_j , X_j , m_j , y_j )$, which includes the following information: $t_j$ is the posting time of the tweet, $X_j$ is the text message, $m_j$ is the topic category and $y_j$ is the stance of the tweet towards a topic or statement. In particular, we consider rumour stance classification where $y_j \epsilon Y$= \textit{\{supporting, denying, questioning, commenting\}}. The stance classification task is to classify the tweet $d_j$ to one of the four classes $y_j \epsilon Y$. 

\section{Background}

\subsection{Point Process}
A point process is random process which models a occurrence of set of points on a real line. If the point process models the occurrence of events over a time period then it is called a temporal point process. For e.g. a point process can be used to model occurrence of earthquakes, rains, etc. A point process can be characterized by its conditional intensity function defined as - 

\begin{equation}
    \lambda(t|\mathcal{H}_t) = \lim_{h \to 0}\dfrac{P(N_{t+h} - N_{t}) = 1|\mathcal{H}_t}{h} 
\end{equation}
where $\mathcal{H}_t$ is history of the process up to time t, with the list of events as $\{t_1, t_2, ... t_n \}$
For the ease of notation, we will denote $ \lambda(t|\mathcal{H}_t)$ as $\lambda(t)$. 
Some of the varieties of commonly used point process are Poisson process, Cox process, Hawkes process etc.

\begin{figure}[t]
    \centering
\includegraphics[width=0.40\textwidth, height=0.20\textwidth]{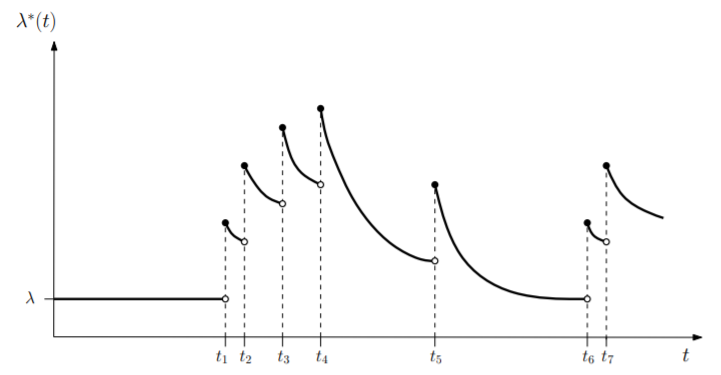}
\caption{Realization of intensity function for Hawkes process}
\label{fig:int_hp}
\end{figure}
\subsection{Hawkes Process}
Point processes are useful to model the distribution of points over some space and are defined using an underlying intensity function.  
A Hawkes process~\citep{hawkes1971spectra} is a point process with self-triggering property i.e occurrence of previous events trigger occurrences of future events. 
Conditional intensity function for univariate Hawkes process is defined as
 \[ \lambda(t) = \mu + \sum_{t_k < t}k(t - t_k) \]
 where 
 $\mu$ is the base intensity function and $k(\cdot)$ is the triggering kernel function capturing the influence from previous events. The summation over $t_k < t$ represents all the effect of all events prior to time $t$ which will contribute in computing the intensity at time $t$. 
Figure~\ref{fig:int_hp} displays the intensity function for Hawkes process which exhibits self-exciting behavior. 
The intensity function for Hawkes process can be 
 Hawkes process has been used in earthquake modelling~\citep{hainzl2010seismicity},crime forecasting~\citep{mohler2011self} and epidemic forecasting~\citep{diggle2005point}. 

\subsection{Multivariate Hawkes Process}
Events are most often associated with features other than time such as categories or users in LBSNs. Such features are known as marks.
The multi-variate Hawkes process~\citep{liniger2009multivariate} is a multi-dimensional point process that can model time-stamped events with marks. It allows explicit representation of marks through the $i^{th}$ dimension of the intensity function and can capture influences across these marks. The intensity function associated with the $i^{th}$ mark is 
\begin{displaymath}
	\lambda_i(t) = \mu_i + \sum_{t_k < t} \alpha_{ii_k} k(t - t_k)
\end{displaymath}
where $\mu_i> 0$ is the base intensity of $i^{th}$ mark. We consider that previous event $k$ is associated with a mark $(i_k)$ and is treated as a dimension in Hawkes process. The intensity at time $t$ for a mark $i$ is assumed to be influenced by all the events happening before $t$ at time $t_k$ and mark $i_k$. The influence of mark $i_k$ on some mark $i$ is given by $\alpha_{ii_k}$. This models the mutual excitation property between events with different marks.

\begin{figure*}[ht]
\begin{subfigure}{.33\textwidth}
\centering
\includegraphics[width=.9\linewidth]{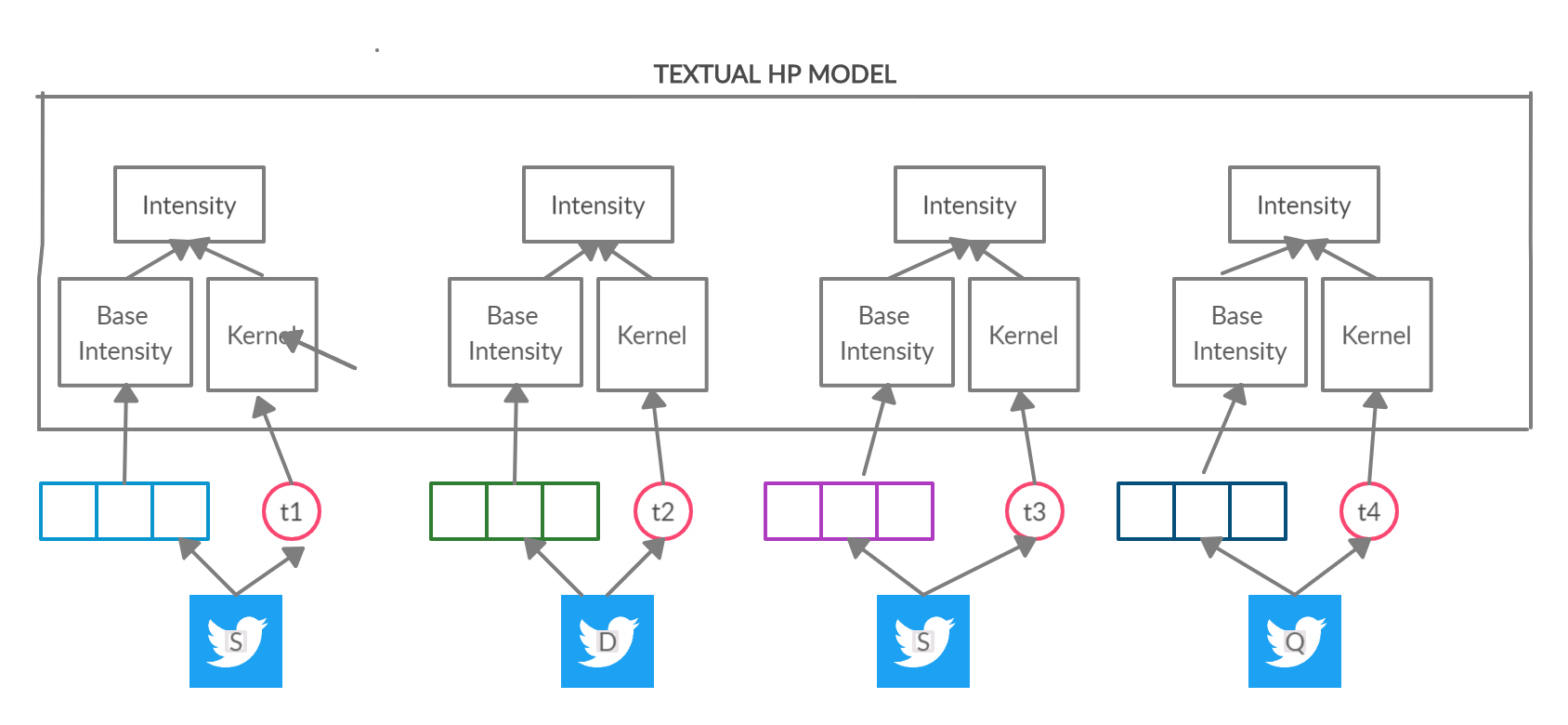}
\caption{Textual HP Model}
\label{fig:framework_textual_hp}
\end{subfigure}
\hfill
\begin{subfigure}{.33\textwidth}
\centering
\includegraphics[width=.9\linewidth]{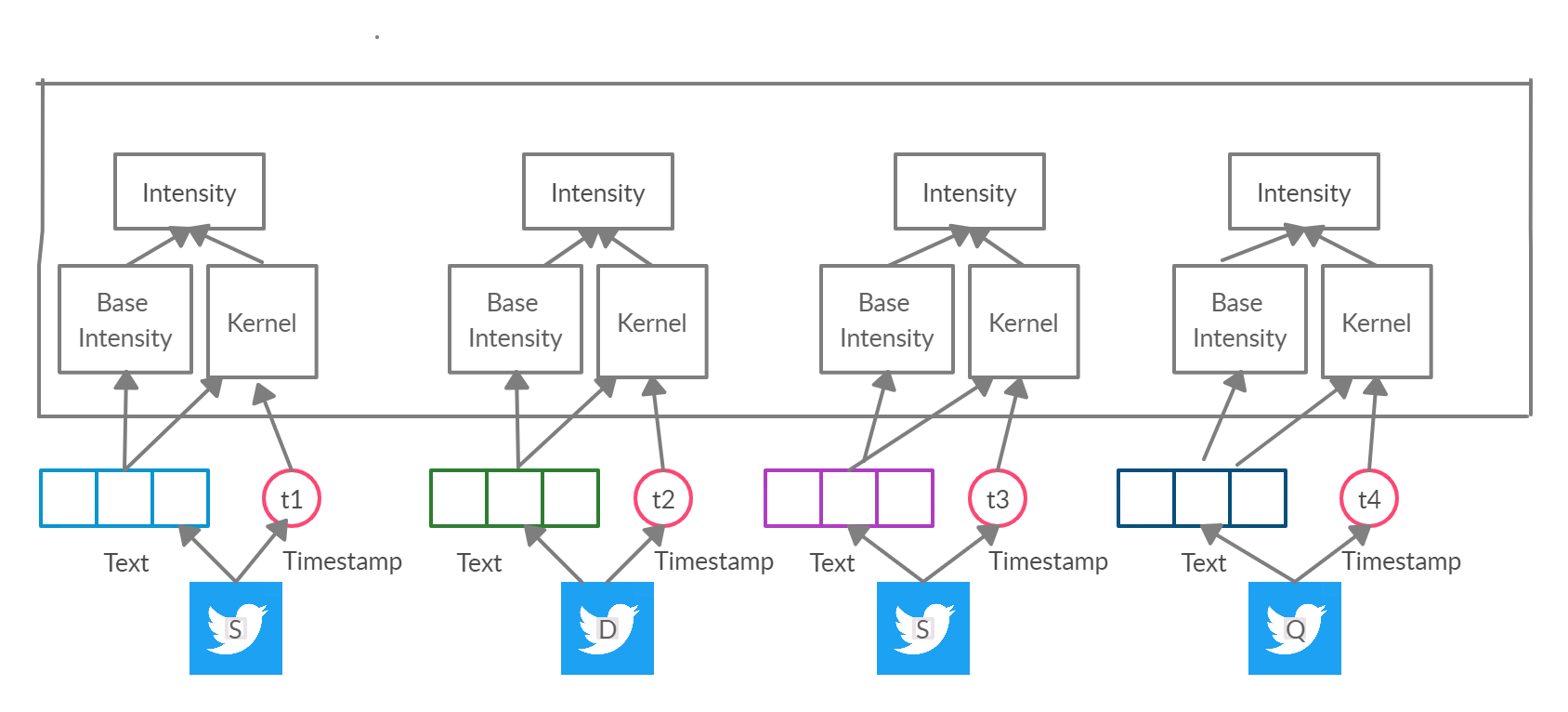}
\caption{Fully Textual HP Model}
\label{fig:framework_fully_textual_hp}
\end{subfigure}
\hfill
\begin{subfigure}{.33\textwidth}
\centering
\includegraphics[width=.9\linewidth]{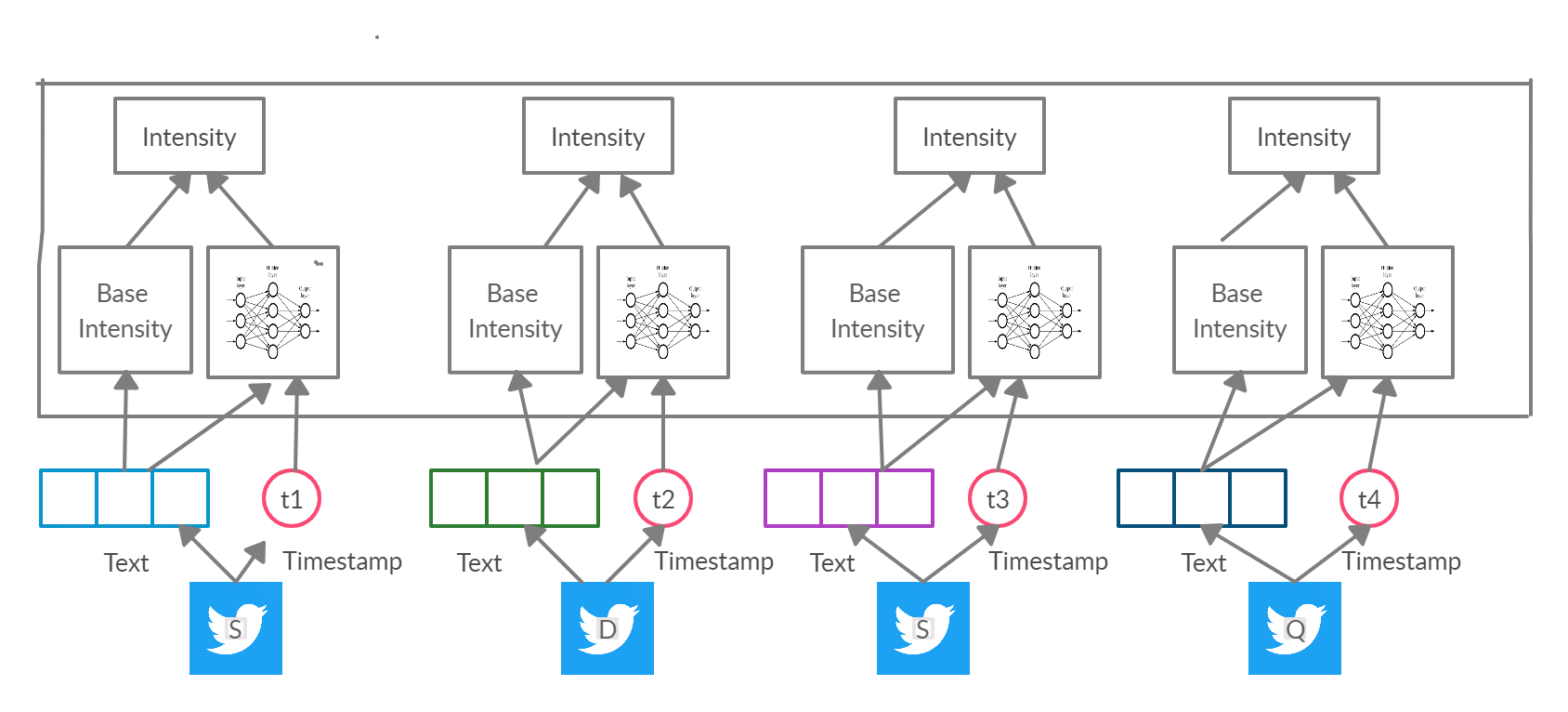}
\caption{Neural Kernel Hawkes Process}
\label{fig:framework_neural_kernel}
\end{subfigure}
    \caption{Framework for the proposed models. \ref{fig:framework_textual_hp})  displays text being used as a part of base intensity \ref{fig:framework_fully_textual_hp})  displays text being used for base intensity as well as kernel
    \label{fig:framework_neural_kernel}) displays kernel is modeled as neural network and text contributes to both base intensity and neural kernel function.}
\end{figure*}
\section{Discriminative Modeling of Text}
Motivated by ~\cite{lukasik2016hawkes}, we also employ a multivariate Hawkes process (MHP) to solve rumour stance classification problem. We treat each dimension of the MHP to correspond to the stance associated with the tweet. The idea is that stances can influence each other and their influence can be captured through the influence matrix of the MHP. 
The intensity function is given by 
\begin{equation}
\label{eq:i1}
    \lambda_{y,m}(t|H_t^-) = \mu_y  + \sum_{t_\ell < t} \mathbf{I}{(m_\ell==m)}  \alpha_{y_\ell,y} k(t-t_\ell)
\end{equation}
Where, the base intensity  $\mu_y$ is a constant base value per stance label and the triggering kernel $k(t-t_\ell) = \omega e^{-\omega(t-t_\ell)}$ (exponentially decaying kernel or excitation function) captures the extent of influence from the past events ($H_t^-$).
The matrix $\alpha$ of size $|Y|\times|Y|$ captures the influence between various classes, e.g. a tweet belonging to class `Support' may have less influence on future tweets belonging to class `Deny' but higher influence on future tweets belonging to class `Support' or `Comment'. But the influence of a 'Support' tweet on a future 'Support' tweet can be low if the future tweets are happening far ahead in time. This will be captured by multiplying the influence matrix with the exponentially decaying triggering kernel, which captures the exponential decay of influence past events on future events.

The likelihood function is given by - 
\begin{equation}
\label{eqn:main_likelihood}
L(t, y, m) = \prod_{n=1}^N  \lambda_{y_n,m_n}(t_n) * \exp(-\sum_{y=1}^{|Y|}\sum_{m=1}^M \int_{0}^T \lambda_{y,m}(s)) ds
\end{equation}
The first part of the likelihood function is the joint likelihood of tweets at time $t_1, ... , t_n$ and the second part provides the likelihood that no tweets happen in the interval [0,T] except at times $t_1 , ... , t_n$. Here, the intensity function can be defined in different ways, depending on the problem set-up.  

For incorporating textual information of tweets, we propose different models where the  textual features are modeled in a discriminative way as a part of intensity function of multivariate Hawkes process model. We present different ways to model text through the base intensity as well as triggering kernel. 
Along with time based kernels, text based kernels can represent the impact of historical events better. We also introduce a methodology where we use a neural network, which is a universal function approximator, as a kernel to model text and time. We discuss the proposed models in detail in the following section:

\subsection{Textual HP: Modeling base intensity using textual features}
\label{section:textual_hp}
The base intensity influences the arrival of events due to exogenous factors. In a standard Hawkes process model, base intensity is constant and learnt from the data. However, we propose a model (\textit{Textual HP}) where base intensity considers the textual features. Along with this, we capture the influence from previous tweets using excitation kernel. In this way, we can model text within the framework of Hawkes process.

\subsubsection{Intensity Function}
Textual HP considers the intensity function to be sum of a base intensity and the excitation function. And base intensity in the proposed model considers the textual features. The base intensity is given by: 
\begin{equation}
\label{bi}
    \mu_{y, t} = \frac{\exp({W_{y}} \times {X_t})}{\sum_{i=1}^{|Y|}\exp({W_{i}} \times {X_t})}
\end{equation}

The base intensity is no longer a constant and depends on the textual content of the post at time $t$. The base intensity is normalized across all labels to avoid it from have a dominating influence on the intensity function.
\begin{itemize} 
\item ${X_t}$ is V-dimensional text representation of tweet at time t.
\item $W$ of size ${|Y|\times |V|}$ are the weights associated with the classes.
\end{itemize}

Using this base intensity, we can write intensity function as:
\begin{equation}
\label{intensity:hptextual}
	\lambda_{y,m}(t|H_t^-) = \mu_{y, t} + \sum_{t_\ell < t} \mathbf{I}{(m_\ell==m)} {\alpha_{y_\ell, y} } \kappa(t - t_\ell)
\end{equation}
As discussed in equation~\ref{eq:i1}, $\alpha_{y_l, y}$ of size ${|Y|\times |Y|}$ captures the influence between various classes of tweets. And $\kappa(t - t_\ell) = \omega \exp(-\omega(t-t_\ell)$ is the exponentially decaying kernel which captures the effect of previous tweets. We can observe that base intensity will be higher for posts whose textual content resembles that of a stance label. Consequently it favours posts with this particular stance if the influence of past labels weighted by time are also favourable.
In this way, we model augment Hawkes process with both time and text based information. 

\subsubsection{Likelihood function}
\label{section:lik_base_hp}
The likelihood function is given by - 
\begin{equation}
L(t, y, X, m) = \big( \prod_{n=1}^N  \lambda_{y_n,m_n}(t_n) \big) * \exp(-\sum_{y=1}^{|Y|}\sum_{m=1}^M \int_{0}^T \lambda_{y,m}(s)
\end{equation}
where the intensity function is defined as in (\ref{intensity:hptextual}). Please note that this is different than equation~\ref{eqn:main_likelihood} in the way that the base intensity used in the intensity function is capable of modeling text as well. The parameters of the  model i.e. $\alpha$ and $W$ are learnt by maximizing the full likelihood.

The log-likelihood function can be given as follows: 
\begin{equation}\label{eq:ll_x}
LL(t, y, X, m) = -\sum_{y=1}^{|Y|} \sum_{m=1}^{M}  \int_{0}^{T} \lambda_{y,m}(s)ds + \sum_{n=1}^N \log \lambda_{y_n,m_n}(t_n)
\end{equation}
We add a regularization term over the weights of text where C is a hyper-parameter for better generalization of model. 
After expanding individual components of equation~(\ref{eq:ll_x}), we get
\begin{equation}\label{final_ll}
    \begin{aligned}
	    LL(t, y, X, m) = \sum_{n=1}^N \log \lambda_{y_n,m_n}(t_n) - |R| \times \sum_{n=1}^{N+1} (t_n - t_{n-1}) \\
	    - \sum_{y=1}^{|Y|} \sum_{l=1}^{N}  \alpha_{y_\ell, y} K(T-t_l) - 
	    C{||W||}^2
	\end{aligned}
\end{equation}
where $K(T-t_l) = 1-exp(-\omega (T-t_\ell))$


\subsubsection{Gradients}
We estimate the parameters by maximizing the log-likelihood  function in equation \ref{final_ll}. We find parameters using joint gradient based optimization over $\alpha$ and $W$, using partial derivatives of log-likelihood. In optimization, we employ L-BFGS approach to gradient search. 
The partial derivatives after expanding equation \ref{final_ll} are given as:
\begin{multline*}
	    \frac{\partial LL}{\partial \alpha_{a,b}} = \sum_{n=1}^{N} (\frac{\mathbf{I}{(y_n==b)} \sum_{l=1}^{n-1} \mathbf{I}{(y_l==a)} \kappa(t-t_l) }{\lambda_{y_n,m_n}(t_n|H_{t_n}^-)}) \\ -\sum_{y=1}^{|Y|} \sum_{l=1}^{N}  \mathbf{I}{(y==b)} \mathbf{I}{(y_l==a)} K(T-t_l)
\end{multline*}
where $K(T-t_\ell) = 1-exp(-\omega (T-t_\ell))$ arises from the integration of $\kappa(t - t_\ell)$. 
\begin{eqnarray}
\label{derivW}
	    \frac{\partial LL}{\partial W_{a,b}} &=& \sum_{n=1}^{N} (\frac{ \exp({W_{y_n}} \times {X_{t_n}}) X_{t_nb} [ \mathbf{I}{(y_n==a)} E - \exp({W_{a}} \times {X_{t_n}}) ] }
	    {E^2 \times \lambda_{y_n,m_n}(t_n|H_{t_n}^-)}) \nonumber \\
	    & & \qquad - 2C  W_{a,b}
\end{eqnarray}
where, $E = \sum_{y=1}^{|Y|}\exp(W_{y} \times {X_{t_n}})$.

\subsection{Fully Textual HP: Using text-based kernel}
We propose a model (\textit{Fully Textual HP}) where we use a text based kernel in combination with the exponential decaying kernel based on time in addition our \textit{Base HP} model discussed in Section~\ref{section:textual_hp}. The text based kernel can help in representing the influence of past events/tweets based on their textual similarity. 

\subsubsection{Intensity Function}
In this case, our model will comprise of text-based base intensity, an exponentially decaying kernel to model time of tweets and a text-based kernel as well.  We can use different types of text kernels like the gaussian kernel, linear kernel, polynomial kernel etc.

Similar to equation~\ref{intensity:hptextual}, we can write the intensity function for the proposed model as:
\begin{equation}
	\lambda_{y,m}(t|H_t^-) = \mu_{y, t} + \sum_{t_\ell < t}  \mathbf{I}{(m_\ell==m)} {\alpha_{y_\ell, y} } \kappa(t - t_\ell) \kappa({X_t}, {X_\ell})
\end{equation}
The base intensity $\mu_{y, t}$ is same as the one used in equation~(\ref{bi}).
The gaussian kernel can be given by -
$$
\kappa({X_t}, {X_\ell}) = \exp{(-\frac{||{X_t} - {X_\ell}||^2}{2*\sigma^2})}
$$
where $\sigma$ is the hyper-parameter. When the textual contents of the post at time $t$ ($X_t$) is similar to a past post text ($X_\ell$) then $\kappa({X_t}, {X_\ell})$ will be higher and consequently the influence of the corresponding label will be higher.  
So, we supplement our intensity function using text using text-based base intensity and kernel. Similar to Section~\ref{section:lik_base_hp}, we can define the likelihood for this model as well.

\subsection{Neural Kernel Hawkes process}
A restriction with the previous approaches is that the past influence is specified through a predefined exponentially decaying kernel function.  Often these influences can take a form other than exponential decay and we intend to capture the functional form of the influence (kernel) through the proposed Neural kernel Hawkes process. In the proposed model, we model kernels using a neural network which is theoretically capable of modelling any function (universal approximator). This is to learn the complex non-linear relationships between historical events and current event.  This is different from the previous works~\cite{mei2017neural,du2016recurrent,xiao2017modeling} combining neural networks with Hawkes process, where the full intensity function is modelled using a neural network losing interpretability advantage of  HP models. Also as we are just modeling kernels using neural networks, we continue to maintain the advantage of explainability of Hawkes process in the form of label-label influence provided  by the $\alpha$ matrix.

\subsubsection{Intensity Function:}
Since the existing kernels can only learn predefined functions, we get extra advantage with neural networks that it can model any function. This model enable us to learn a more generalized version of Hawkes process keeping its causality intact. The intensity function can be defined as:
\begin{equation}
    \lambda_{y,m}(t) = \mu_{y, t} + \sum_{t_\ell < t} \mathbf{I}{(m_\ell,m)} {\alpha_{y_\ell, y} } F([t_\ell, X_\ell], [t, {X_t}]; W_{nn}) 
    \end{equation}
where $W_{nn}$ are the weights in the NN kernel, the text and time are input together and the base intensity $\mu_{y, t}$ is defined in (\ref{bi}). 

All the parameters including the neural network kernel parameters are learnt by maximizing the likelihood defined in (\ref{eq:ll_x}). We approximate the intractable integral using the Monte Carlo approximation.  It  computes average intensity over uniformly sampled time and multiply with time period to get  integral value. Backpropagation is applied on the (\ref{eq:ll_x}) to learn parameters of neural kernel.
Prediction is done by evaluating the intensity function across all the classes at the time of the post and choosing the class with the highest intensity function.

\begin{figure}[b]
    \centering
    \abovecaptionskip=1pt 
\belowcaptionskip=0.5pt 
    \includegraphics[width=0.50\textwidth]{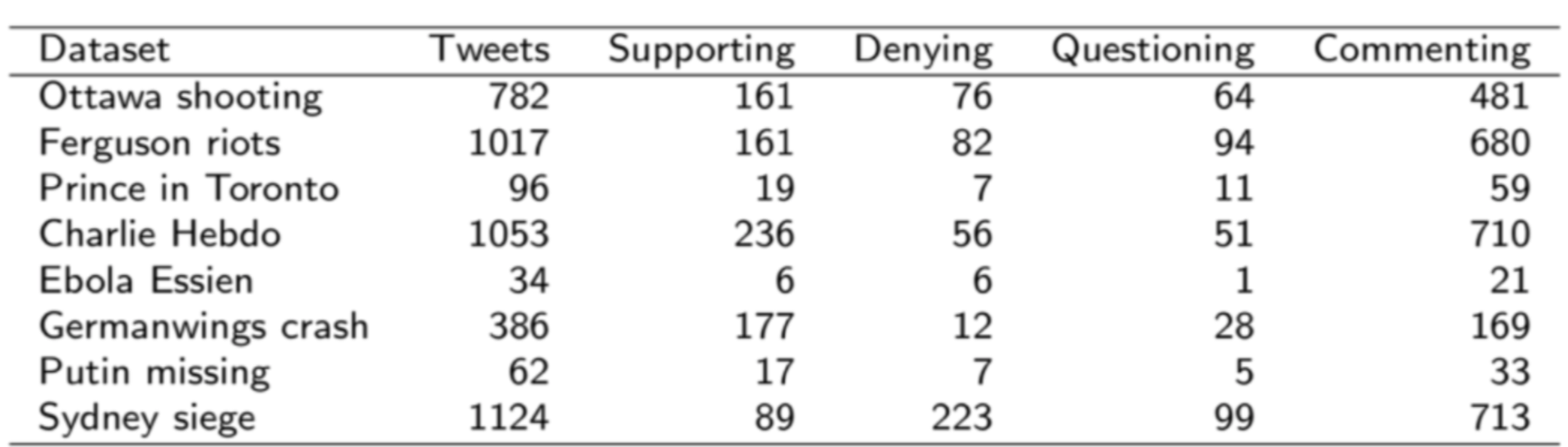}
    \caption{Statistics of the PHEME dataset}
    \label{fig:dataset}
\end{figure}

\section{Experiments}
\subsection{Dataset}
We use the PHEME dataset~\cite{zubiaga2016analysing} for rumour stance classification. It considers tweets belonging to nine noteworthy events occurred around the world. Along with tweets, it also considers retweets, and replies to form a tweet thread.   The dataset contains a set of rumour threads. Each thread contains a source tweet as well as replies to that tweet. Every tweet is assigned a stance of - Supporting, Denying, Questioning, Commenting classes w.r.t. the source tweet. 
The detailed statistics of the dataset used is mentioned in Figure~\ref{fig:dataset}. One notable characteristic of the dataset is that the distribution of categories is skewed towards commenting tweets, and that this imbalance varies slightly across the eight events. This varying imbalance makes the task more realistic and challenging.

\subsection{Baselines}
We have considered following baselines:
\begin{itemize}
    \item \textbf{Hawkes Process:~\cite{lukasik2016hawkes}} The authors have considered two approaches for optimization using gradient based optimization and approximating the parameters. They have been shown to perform better than several machine learning models including conditional random fields. We have compared our results with both the approaches used. 
    \item \textbf{LSTM:~\cite{zubiaga2018discourse}} The authors have used the sequential structure of conversational threads using LSTM. 
\end{itemize}



\begin{figure}[b]
    \centering
    \includegraphics[width=0.40\textwidth]{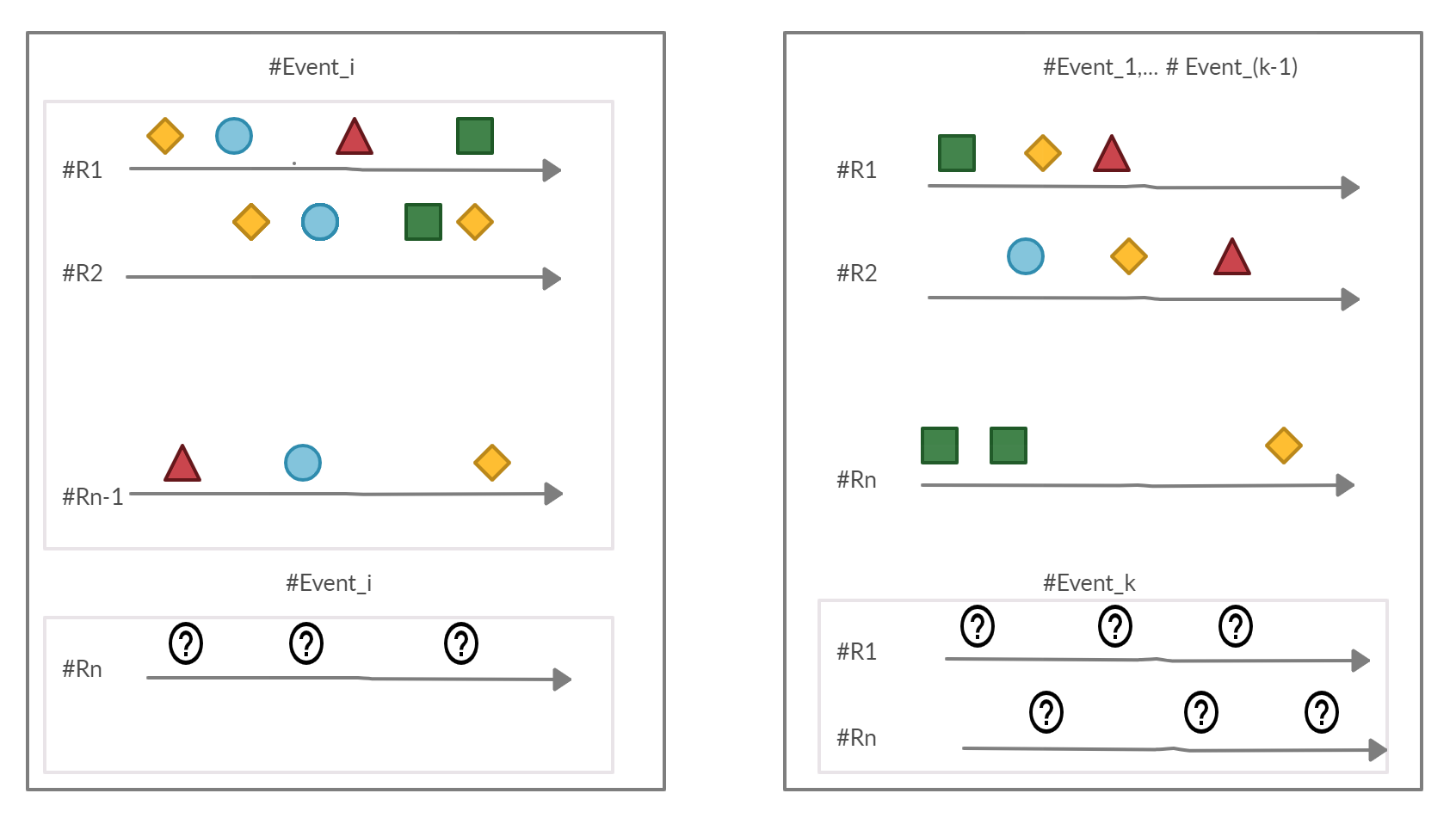}
    \caption{Experimental set-up (left) shows Leave one out - Thread and (right) shows Leave one out - Event} 
    \label{fig:exp_setup}
\end{figure}

\begin{table*}[t]
\centering
 \caption{Results from the baselines (bottom rows) and our proposed approaches (top rows). First four results columns are from individual four rumor datasets, and the last column corresponds to the leave one event out, where each rumor event is held out for prediction.}
\label{tab:R1}
\begin{tabular}{lcccccccc} 
\toprule
                            & \multicolumn{2}{c}{Ottawa}       & \multicolumn{2}{c}{Ferguson}      & \multicolumn{2}{c}{Charlie Hebdo} & \multicolumn{2}{c}{Sydney Siege}      \\ 
                            & Acc.(\%)        & F1              & Acc.(\%)         & F1              & Acc.(\%)         & F1              & Acc.(\%)         & F1                         \\ 
\midrule
\multicolumn{1}{l}{Base Textual HP}                  & \textbf{70.2}  & 0.312           & \textbf{71.12}  & 0.329  & 69.5            & 0.324           & \textbf{71.63}  & 0.324                 \\
\multicolumn{1}{l}{Fully textual HP}        & 62.4           & 0.329           & 62.34           & 0.259           & 69.04           & 0.304           & 64.32           & 0.318             \\
\multicolumn{1}{l}{Neural Kernel HP}  & 56.01          & 0.153           & 66.7            & 0.193           & 59.18           & 0.169           & 56.32           & 0.168                     \\ 
\midrule
\multicolumn{1}{l}{HP Grad~\cite{lukasik2016hawkes}}    & 63.43          & 0.424           & 63.23           & 0.331           & 71.79           & 0.419           & 62.99           & 0.395          \\
\multicolumn{1}{l}{HP Approx. ~\cite{lukasik2016hawkes}} & 67.77          & 0.32            & 68.44           & 0.26            & \textbf{72.93}  & 0.325           & 68.59           & 0.349                  \\
\multicolumn{1}{l}{LSTM on HPfeatures~\cite{zubiaga2018discourse}}          & 66.67          & \textbf{0.487}  & 69.73           & \textbf{0.409}          & 70.99           & \textbf{0.513}  & 69.51           & \textbf{0.496}                         \\
\bottomrule
\end{tabular}
\end{table*}
\subsection{Experimental Setup}
The experiments are tried to build in a way where we depict real world scenarios as closely as possible. In real world, new rumours arise on a regular basis.  We  train models on old rumours and then use them for stance classification on new rumours. We try to perform something similar in our experiments. We call it the 'leave one out' approach. We train on a set of rumours and then test on a new unseen rumour. The experimental setup  can be categorized into two types.
\paragraph{\textbf{Leave one out - Thread}}\label{loot}
Following prior work~\cite{DBLP:journals/corr/LukasikBCZLP16}, we consider 4 events - Ottawa, Ferguson, Charlie Hebdo and Sydney Siege, the largest events from PHEME (each with approximately $1000$ tweets per event). 
Every event in the data set has multiple tweet threads ($50-70$), where each thread is a new rumour generated when the event occurred. We train on $n-1$ rumour threads and test on the $n^{th}$ rumour. We perform this $n$ times, testing on a different rumour each time. This helps in getting the overall performance across all rumours.
\paragraph{\textbf{Leave one out - Event}}\label{looe}
Here, a dataset of top 8 events is considered and then combined to form a bigger data set, with $4554$ tweets in total. 
We consider training on 7 events at a time and testing on the $8^{th}$ one. This is repeated 8 times, and an average score is reported.

\paragraph{\textbf{Evaluation Metrics}}
We use the popular metrics for multi-class classification i.e. accuracy and F1 scores. We consider micro-averaged accuracy and   macro averaged F1 score as reported in the previous work. 
Macro-averaged F1-score can be calculated as the harmonic mean of  macro-averaged precision and recall.
Considering number of stances to be $K$, the formulae for macro F1-score can be written as follows - 
\begin{equation*}
\begin{split}
    & Macro\text{-}Precision = \frac{\sum_{i=1}^K Precision}{K} \\
    & Macro\text{-}Recall = \frac{\sum_{i=1}^K Recall}{K} \\
    & Macro\text{-}F1 Score = \frac{2*Macro\text{-}Precision*Macro\text{-}Recall} {Macro\text{-}Precision + Macro\text{-}Recall}
\end{split}
\end{equation*}
\subsection{Implementation Details}
We have performed our experiments on Intel(R) Xeon(R) processor with 2.70GHz CPU and 125 GB RAM. Neural Kernel HP Model has been run on Tesla P-100 infrastructure. The tweets are subjected to various pre-processing steps like removal of stopwords, URLs and punctuations; replacing emoticons, user mentions and URLs followed by stemming. After preprocessing, we use standard 100-dimensional word2vec (trained on Google News) representation obtained by averaging the word2vec representation of each token in a tweet. The best configuration for Base HP Textual is for regularization parameter and temporal kernel parameter as 0.05. For Fully HP Textual, best results have been achieved for regularization parameter, temporal kernel parameter and text kernel parameter as 0.05. Best configuration for Neural Network Kernel model includes 2 hidden layers with 20 neurons each for 0.005 learning rate and 0.9 momentum. The optimizer used in AdaGrad. And 50 samples have been used for Monte Carlo approximation.

\section{Results and Analysis}
\subsection{Results}
\begin{figure}[t]
    \centering
    \abovecaptionskip=1pt 
\belowcaptionskip=0.5pt 
    \includegraphics[width=0.40\textwidth]{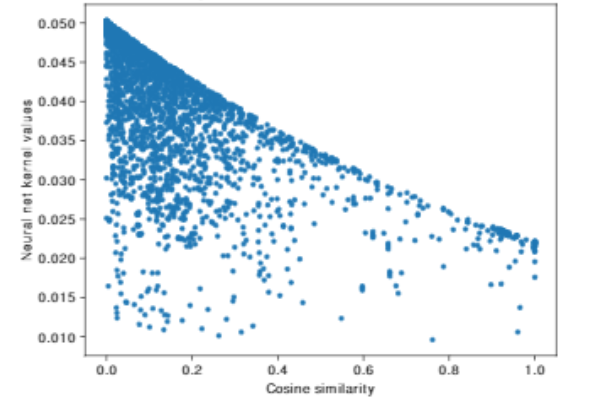}
    \caption{Neural kernel values for text features of an input pair of tweets vs. cosine similarity between these text features. Dots accumulated over the time difference range 0-1 hours.} 
    \label{fig:nn_text}
\end{figure}

The proposed approach is compared against the Hawkes process~\cite{lukasik2016hawkes} and LSTM~\cite{zubiaga2018discourse} based approaches for rumour stance classification. The \textit{Base Textual HP} which used discriminative modeling of text with normalized base intensity outperforms the benchmarks for all events except for Charlie Hebdo in terms of micro-accuracy, demonstrating that including textual features as part of intensity function helps improve results. In comparison with the LSTM approach ~\cite{zubiaga2018discourse} for this setup, we can see that our Textual HP model gives better accuracy in all datasets except Charlie Hebdo. This shows usefulness of HP based models over modern neural networks especially when dataset size is small.
The \textit{Fully textual HP} which uses text in base intensity as well as kernel, gives comparable results to benchmark models, but doesn't outperform \textit{Base Textual HP}. This means that influence arising through text similarity is not very useful for predictions at thread level, with typical thread size being 10. Here, dissimilar text belonging to different classes (e.g. deny and support tweets) can have higher influence, which is restricted through the text kernel. Although, this shows another successful way of augmenting Hawkes process with text.
 We also observe that \textit{Neural Kernel HP} did not give good performance.
In Figure~\ref{fig:nn_text} we show an example function learned by the neural kernel against text similarity and in general, and we find a decrease w.r.t cosine similarity. 
This supports the observations from fully textual HP. 
However, Neural kernel HP did not perform well overall, presumably due to the small sized rumour stance data (1000 tweets per event). 
\begin{table}[t]
\centering
\caption{Result comparison in Leave one out - Event setup}
\label{tab:R2}
\begin{tabular}{@{}lcc@{}}
\toprule
                        & Accuracy (\%) & Macro F1 \\ \midrule
Textual HP      & 64.70         & 0.269    \\
Fully textual HP     & \textbf{69.10}         & \textbf{0.329}    \\
Neural kernel HP     & 59.44         & 0.233    \\
HP Grad. \cite{lukasik2016hawkes}   & -             & 0.309    \\
HP Approx. \cite{lukasik2016hawkes} & -             & 0.307    \\
LSTM on Hawkes Features \cite{zubiaga2018discourse} & -             & 0.318    \\ \bottomrule
\end{tabular}
\end{table}

We can see the results for Leave one out - event approach explained in Section \ref{looe} in Table \ref{tab:R2}.
The \textit{Fully Textual HP} gives the best results beating the benchmarks, showing importance of considering text similarities between posts, as opposed to only similarities between categories from~\cite{lukasik2016hawkes}.
The \textit{Neural Kernel HP} model performs better in this setup, however is still limited by the small data set size. 
On the other hand, using the inductive bias of Hawkes process assumption helps perform better under this data scarce scenario.

 \begin{table}[b]
 \centering
 \tabcolsep=0.2cm
 \caption{Influence matrix of Fully Textual HP model results}\label{Tab:a2}
 \begin{tabular}{lllll}
 \hline
                           & \textit{\textbf{Support}} & \textit{\textbf{Deny}} & \textit{\textbf{Question}} & \textit{\textbf{Comment}} \\ \hline
 \textit{\textbf{Support}}  & \textit{0.0119}                    & 0.0087                 & 0.0062                     & \textbf{0.1643}                    \\ 
 \textit{\textbf{Deny}}     & 0.0129                   & \textit{0.0135}                 & \textbf{0.0136}                     & 0.0127                    \\ 
 \textit{\textbf{Question}} & 0.0146                    & 0.0128                 & \textit{0.0149}                    & \textbf{0.1003}                    \\ 
 \textit{\textbf{Comment}}  & \textit{0.0043}                    & 0.0022                 & 0.0013                     & \textbf{0.0559}                    \\ 
 \bottomrule
 \end{tabular}
 \end{table}
 
\subsection{Analysis}
\subsubsection{Analysis of Influence Matrix $\alpha$}
We analyze the values learnt by the influence matrix, $\alpha$. It is a $4 \times 4$ dimensional matrix which learns the influence of different classes of tweets on others. For example it learns the impact of a previous tweet being of class Support on next tweet being of class Deny. In Table~\ref{Tab:a2}, we see sample values of influence matrix belonging to Fully Textual HP model. The bold face values show the best result in a row while italics show the second best. An interesting observation is that Deny class has highest influence on the Question and Deny classes. This means that a deny tweet is often followed by a question tweet or a Deny tweet, which makes sense in rumour stance classification.  The diagonal values are relatively high which means that each class influences the next tweet to belong to that class, i.e. Support or Question  is likely to attract more Support or Question tweets respectively than tweets belonging to other classes. The values  in the last column are also usually quite high in the row. The last column belongs to Comments class. This tells that it is very likely for a comment tweet to follow tweets belonging to other classes. This also is quite expected as the data set has a class imbalance with more than $60\%$ tweets belonging to Comments class. 

\begin{figure}[t]
    \centering
\abovecaptionskip=1pt 
\belowcaptionskip=0.5pt 
\includegraphics[width=0.40\textwidth]{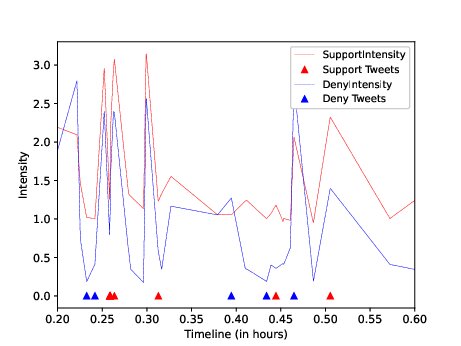}
\caption{A snapshot of posts and intensities for 2 classes, Support and Deny, at post times using  \textit{Base Textual HP} on Sydney Siege data. The intensity for a class becomes higher as some tweet occurs from that class through  textual features and temporal influences}
\label{fig:int_plot}
\end{figure}

\subsubsection{Intensity plots}
Figure \ref{fig:int_plot} plots intensity of Textual HP for \textit{Support} and \textit{Deny} class tweets. 
Tweets are associated with posting times, and in Figure \ref{fig:int_plot} we plot  the intensity value of tweets at their posting times for support and deny classes. The intensity function values are obtained by considering  temporal and textual information as discussed in equation~\ref{eq:i1}.  In Figure \ref{fig:int_plot}, we find that intensity value is higher for the tweets of respective classes.
Figure~\ref{fig:nn_text} shows the kernel function learnt against cosine similarity of text for Neural Kernel.  The pairs of tweets are selected such that they belong to the same thread. We compute the cosine similarity (using text) and the neural kernel values (using text and time)  between them.  The difference in their time of occurrence ranges  between 0-1hrs. Hence there can be multiple pairs with the same cosine similarity. We can observe here that in general neural kernel value is decreasing with increasing cosine similarity.



\begin{figure}[] 
    \centering
    \subfloat[]{
   \includegraphics[width=0.40\textwidth, height=0.40\textwidth,keepaspectratio]{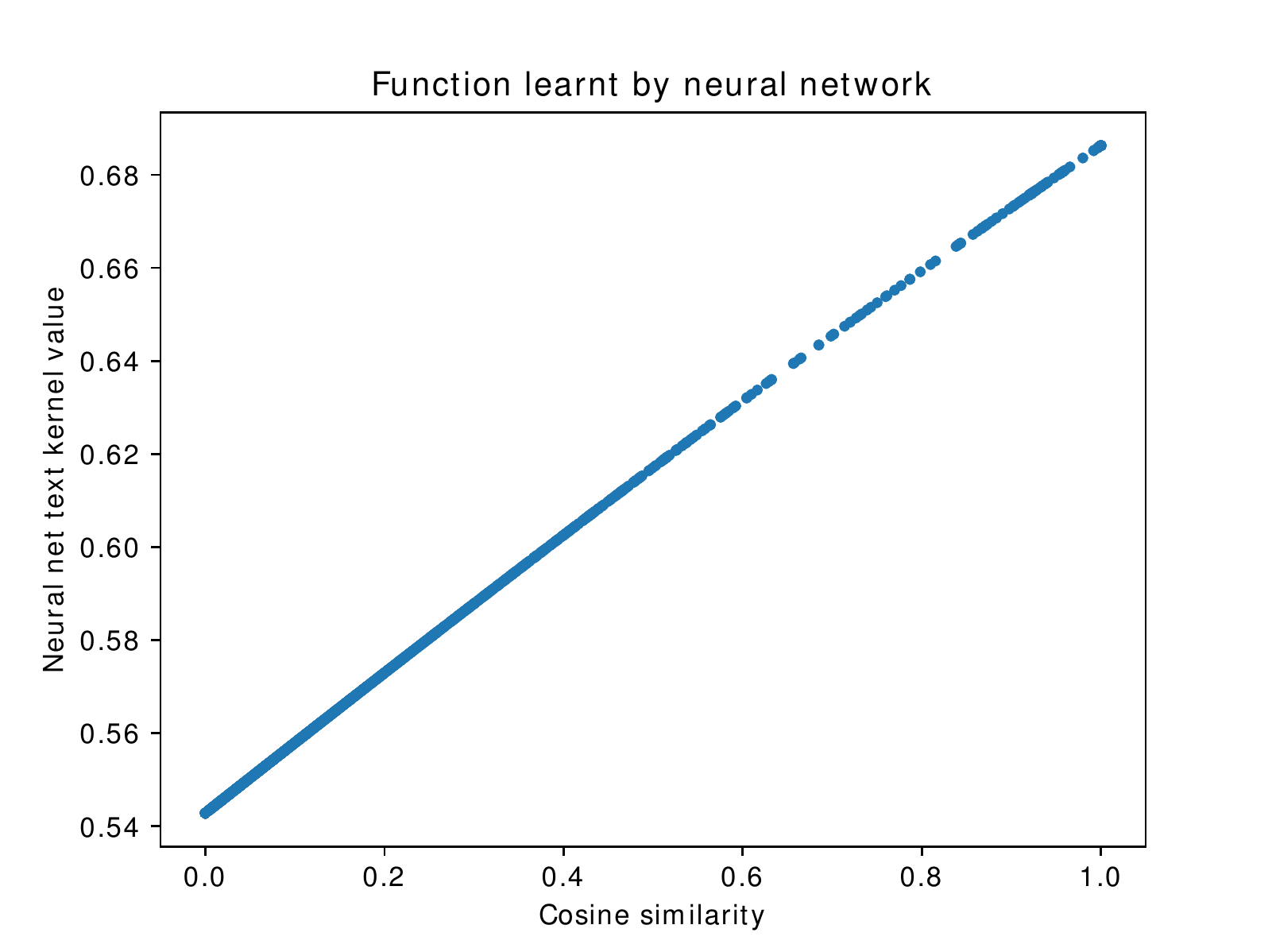}
\label{fig:nn_text_abl}
  }
  
\subfloat[]{
 \includegraphics[width=0.40\textwidth, height=0.40\textwidth,keepaspectratio]{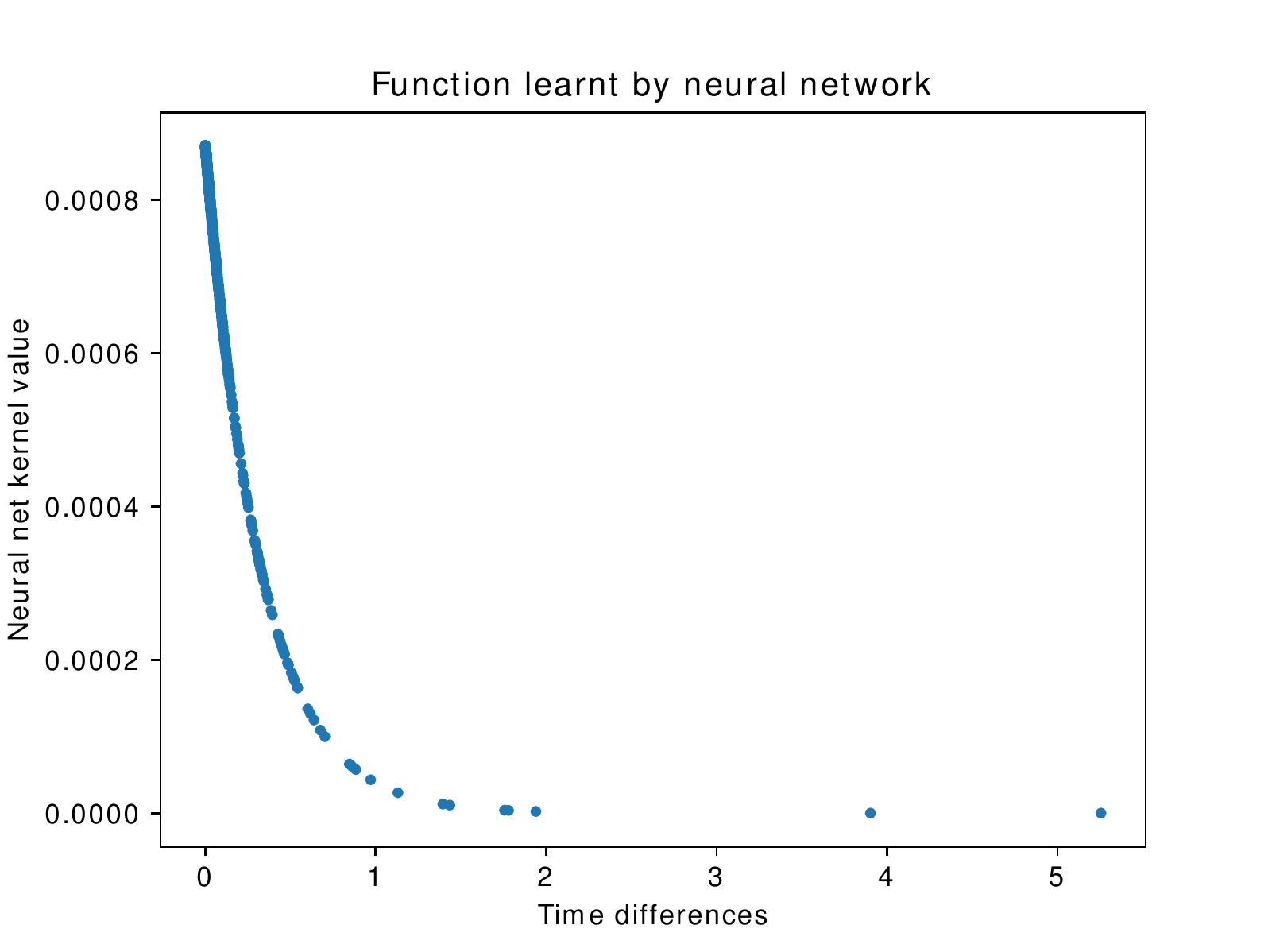}
  \label{fig:nn_time}
  }
    \caption{\ref{fig:nn_text_abl}) displays function learnt by independent text neural network when we consider neural network with two kernels and \ref{fig:nn_time}) displays function learnt for time by neural network kernel}
    \label{fig:nn}
\end{figure}


\subsubsection{Function learnt by neural networks}
Figure \ref{fig:nn_time} shows the function learnt by neural network with a single time-based kernel in Neural Kernel HP. We can see that it learns a kernel similar to exponential decaying kernel. In Figure \ref{fig:nn_text} we can see the function learnt by text-based neural network when we use separate networks for both time and text. Its  an increasing function of cosine similarity. This conveys that we get higher value when the text of historical events are similar to current event.

\section{Ablation Study}
We carry out an ablation study on Neural Hawkes Process model by using text and time based kernels in different ways. Thereby, we learn the nature of the function learnt by neural network. Figure~\ref{fig:nn_time} shows the time kernel learnt when we use input just the difference of time for the neural network. We can observe that it learns a function similar to exponentially decaying kernel, which explains the relevance of exponential kernel for such settings. In another variation, we use two separate kernels for time and text. Figure~\ref{fig:nn_text} displays the function learnt for text under such settings. The value of function learnt increases with increasing similarity of text. However, results of both the variants were not as good as the proposed model. 

\section{Conclusion}
We proposed a novel method for text classification based on Hawkes processes, where we consider textual features, expanding on previous applications of this model. 
In particular, we propose a text based kernel and a neural kernel over text and time, providing more flexible approaches to modeling influences among data points. We propose three methods to model text within the intensity function. Then we perform discriminative modeling and use the intensity to obtain labels. This enables us to capture the influence among tweets not only using time but also using  textual content of tweets. We propose using  kernels (exponential and neural network) over text and time for modelling influences. Neural network kernel can learn the functional form of the influence rather than predefining as exponential function. This allows to easily consider pre-trained word embeddings in the model for effective text classification. The experiments on rumour stance classification showed the effectiveness of the proposed approaches. 
We infer from the results that the Textual HP model outperforms generative models of text with HP. We also show that HP based approaches can perform better than neural networks on smaller datasets. It opens up a new direction to use text within the framework of Hawkes process.

\bibliographystyle{ACM-Reference-Format}
\bibliography{ArxivPaper}


\begin{thebibliography}{27}


\ifx \showCODEN    \undefined \def \showCODEN     #1{\unskip}     \fi
\ifx \showDOI      \undefined \def \showDOI       #1{#1}\fi
\ifx \showISBNx    \undefined \def \showISBNx     #1{\unskip}     \fi
\ifx \showISBNxiii \undefined \def \showISBNxiii  #1{\unskip}     \fi
\ifx \showISSN     \undefined \def \showISSN      #1{\unskip}     \fi
\ifx \showLCCN     \undefined \def \showLCCN      #1{\unskip}     \fi
\ifx \shownote     \undefined \def \shownote      #1{#1}          \fi
\ifx \showarticletitle \undefined \def \showarticletitle #1{#1}   \fi
\ifx \showURL      \undefined \def \showURL       {\relax}        \fi
\providecommand\bibfield[2]{#2}
\providecommand\bibinfo[2]{#2}
\providecommand\natexlab[1]{#1}
\providecommand\showeprint[2][]{arXiv:#2}

\bibitem[\protect\citeauthoryear{Aker, Derczynski, and Bontcheva}{Aker
  et~al\mbox{.}}{2017}]%
        {aker2017simple}
\bibfield{author}{\bibinfo{person}{Ahmet Aker}, \bibinfo{person}{Leon
  Derczynski}, {and} \bibinfo{person}{Kalina Bontcheva}.}
  \bibinfo{year}{2017}\natexlab{}.
\newblock \showarticletitle{Simple open stance classification for rumour
  analysis}.
\newblock \bibinfo{journal}{\emph{arXiv preprint arXiv:1708.05286}}
  (\bibinfo{year}{2017}).
\newblock


\bibitem[\protect\citeauthoryear{Bedathur, Bhattacharya, Choudhari, and
  Dasgupta}{Bedathur et~al\mbox{.}}{2018}]%
        {bedathur2018discovering}
\bibfield{author}{\bibinfo{person}{Srikanta Bedathur},
  \bibinfo{person}{Indrajit Bhattacharya}, \bibinfo{person}{Jayesh Choudhari},
  {and} \bibinfo{person}{Anirban Dasgupta}.} \bibinfo{year}{2018}\natexlab{}.
\newblock \showarticletitle{Discovering topical interactions in text-based
  cascades using hidden Markov Hawkes processes}.
\newblock \bibinfo{journal}{\emph{arXiv preprint arXiv:1809.04487}}
  (\bibinfo{year}{2018}).
\newblock


\bibitem[\protect\citeauthoryear{Diggle, Rowlingson, and Su}{Diggle
  et~al\mbox{.}}{2005}]%
        {diggle2005point}
\bibfield{author}{\bibinfo{person}{Peter Diggle}, \bibinfo{person}{Barry
  Rowlingson}, {and} \bibinfo{person}{Ting-li Su}.}
  \bibinfo{year}{2005}\natexlab{}.
\newblock \showarticletitle{Point process methodology for on-line
  spatio-temporal disease surveillance}.
\newblock \bibinfo{journal}{\emph{Environmetrics: The official journal of the
  International Environmetrics Society}} \bibinfo{volume}{16},
  \bibinfo{number}{5} (\bibinfo{year}{2005}), \bibinfo{pages}{423--434}.
\newblock


\bibitem[\protect\citeauthoryear{Du, Dai, Trivedi, Upadhyay, Gomez-Rodriguez,
  and Song}{Du et~al\mbox{.}}{2016}]%
        {du2016recurrent}
\bibfield{author}{\bibinfo{person}{Nan Du}, \bibinfo{person}{Hanjun Dai},
  \bibinfo{person}{Rakshit Trivedi}, \bibinfo{person}{Utkarsh Upadhyay},
  \bibinfo{person}{Manuel Gomez-Rodriguez}, {and} \bibinfo{person}{Le Song}.}
  \bibinfo{year}{2016}\natexlab{}.
\newblock \showarticletitle{Recurrent marked temporal point processes:
  Embedding event history to vector}. In \bibinfo{booktitle}{\emph{Proceedings
  of the 22nd ACM SIGKDD International Conference on Knowledge Discovery and
  Data Mining}}. ACM, \bibinfo{pages}{1555--1564}.
\newblock


\bibitem[\protect\citeauthoryear{Du, Farajtabar, Ahmed, Smola, and Song}{Du
  et~al\mbox{.}}{2015}]%
        {du2015dirichlet}
\bibfield{author}{\bibinfo{person}{Nan Du}, \bibinfo{person}{Mehrdad
  Farajtabar}, \bibinfo{person}{Amr Ahmed}, \bibinfo{person}{Alexander~J
  Smola}, {and} \bibinfo{person}{Le Song}.} \bibinfo{year}{2015}\natexlab{}.
\newblock \showarticletitle{Dirichlet-hawkes processes with applications to
  clustering continuous-time document streams}. In
  \bibinfo{booktitle}{\emph{Proceedings of the 21th ACM SIGKDD International
  Conference on Knowledge Discovery and Data Mining}}.
  \bibinfo{pages}{219--228}.
\newblock


\bibitem[\protect\citeauthoryear{{Dutta}, {Dutta}, {Adhikary}, and
  {Chakraborty}}{{Dutta} et~al\mbox{.}}{2020}]%
        {Dutta2020HawkesEye}
\bibfield{author}{\bibinfo{person}{H.~S. {Dutta}}, \bibinfo{person}{V.~R.
  {Dutta}}, \bibinfo{person}{A. {Adhikary}}, {and} \bibinfo{person}{T.
  {Chakraborty}}.} \bibinfo{year}{2020}\natexlab{}.
\newblock \showarticletitle{HawkesEye: Detecting Fake Retweeters Using Hawkes
  Process and Topic Modeling}.
\newblock \bibinfo{journal}{\emph{IEEE Transactions on Information Forensics
  and Security}}  \bibinfo{volume}{15} (\bibinfo{year}{2020}),
  \bibinfo{pages}{2667--2678}.
\newblock


\bibitem[\protect\citeauthoryear{Hainzl, Steacy, and Marsan}{Hainzl
  et~al\mbox{.}}{2010}]%
        {hainzl2010seismicity}
\bibfield{author}{\bibinfo{person}{Sebastian Hainzl}, \bibinfo{person}{D
  Steacy}, {and} \bibinfo{person}{S Marsan}.} \bibinfo{year}{2010}\natexlab{}.
\newblock \showarticletitle{Seismicity models based on Coulomb stress
  calculations}.
\newblock \bibinfo{journal}{\emph{Community Online Resource for Statistical
  Seismicity Analysis}} (\bibinfo{year}{2010}).
\newblock


\bibitem[\protect\citeauthoryear{Hawkes}{Hawkes}{1971}]%
        {hawkes1971spectra}
\bibfield{author}{\bibinfo{person}{Alan~G Hawkes}.}
  \bibinfo{year}{1971}\natexlab{}.
\newblock \showarticletitle{Spectra of some self-exciting and mutually exciting
  point processes}.
\newblock \bibinfo{journal}{\emph{Biometrika}} \bibinfo{volume}{58},
  \bibinfo{number}{1} (\bibinfo{year}{1971}), \bibinfo{pages}{83--90}.
\newblock


\bibitem[\protect\citeauthoryear{Hawkes and Oakes}{Hawkes and Oakes}{1974}]%
        {hawkes1974cluster}
\bibfield{author}{\bibinfo{person}{Alan~G Hawkes} {and} \bibinfo{person}{David
  Oakes}.} \bibinfo{year}{1974}\natexlab{}.
\newblock \showarticletitle{A cluster process representation of a self-exciting
  process}.
\newblock \bibinfo{journal}{\emph{Journal of Applied Probability}}
  \bibinfo{volume}{11}, \bibinfo{number}{3} (\bibinfo{year}{1974}),
  \bibinfo{pages}{493--503}.
\newblock


\bibitem[\protect\citeauthoryear{He, Rekatsinas, Foulds, Getoor, and Liu}{He
  et~al\mbox{.}}{2015}]%
        {he2015Hawkestopic}
\bibfield{author}{\bibinfo{person}{Xinran He}, \bibinfo{person}{Theodoros
  Rekatsinas}, \bibinfo{person}{James Foulds}, \bibinfo{person}{Lise Getoor},
  {and} \bibinfo{person}{Yan Liu}.} \bibinfo{year}{2015}\natexlab{}.
\newblock \showarticletitle{Hawkestopic: A joint model for network inference
  and topic modeling from text-based cascades}. In
  \bibinfo{booktitle}{\emph{International conference on machine learning}}.
  \bibinfo{pages}{871--880}.
\newblock


\bibitem[\protect\citeauthoryear{Jebara and Pentland}{Jebara and
  Pentland}{1999}]%
        {jebara1999maximum}
\bibfield{author}{\bibinfo{person}{Tony Jebara} {and} \bibinfo{person}{Alex
  Pentland}.} \bibinfo{year}{1999}\natexlab{}.
\newblock \showarticletitle{Maximum conditional likelihood via bound
  maximization and the CEM algorithm}. In \bibinfo{booktitle}{\emph{Advances in
  neural information processing systems}}. \bibinfo{pages}{494--500}.
\newblock


\bibitem[\protect\citeauthoryear{Kochkina, Liakata, and Augenstein}{Kochkina
  et~al\mbox{.}}{2017}]%
        {kochkina2017turing}
\bibfield{author}{\bibinfo{person}{Elena Kochkina}, \bibinfo{person}{Maria
  Liakata}, {and} \bibinfo{person}{Isabelle Augenstein}.}
  \bibinfo{year}{2017}\natexlab{}.
\newblock \showarticletitle{Turing at semeval-2017 task 8: Sequential approach
  to rumour stance classification with branch-lstm}.
\newblock \bibinfo{journal}{\emph{arXiv preprint arXiv:1704.07221}}
  (\bibinfo{year}{2017}).
\newblock


\bibitem[\protect\citeauthoryear{Kochkina, Liakata, and Zubiaga}{Kochkina
  et~al\mbox{.}}{2018}]%
        {kochkina2018all}
\bibfield{author}{\bibinfo{person}{Elena Kochkina}, \bibinfo{person}{Maria
  Liakata}, {and} \bibinfo{person}{Arkaitz Zubiaga}.}
  \bibinfo{year}{2018}\natexlab{}.
\newblock \showarticletitle{All-in-one: Multi-task learning for rumour
  verification}.
\newblock \bibinfo{journal}{\emph{arXiv preprint arXiv:1806.03713}}
  (\bibinfo{year}{2018}).
\newblock


\bibitem[\protect\citeauthoryear{Liniger}{Liniger}{2009}]%
        {liniger2009multivariate}
\bibfield{author}{\bibinfo{person}{Thomas~Josef Liniger}.}
  \bibinfo{year}{2009}\natexlab{}.
\newblock \emph{\bibinfo{title}{Multivariate hawkes processes}}.
\newblock \bibinfo{thesistype}{Ph.D. Dissertation}. \bibinfo{school}{ETH
  Zurich}.
\newblock


\bibitem[\protect\citeauthoryear{Lukasik, Bontcheva, Cohn, Zubiaga, Liakata,
  and Procter}{Lukasik et~al\mbox{.}}{2016a}]%
        {DBLP:journals/corr/LukasikBCZLP16}
\bibfield{author}{\bibinfo{person}{Michal Lukasik}, \bibinfo{person}{Kalina
  Bontcheva}, \bibinfo{person}{Trevor Cohn}, \bibinfo{person}{Arkaitz Zubiaga},
  \bibinfo{person}{Maria Liakata}, {and} \bibinfo{person}{Rob Procter}.}
  \bibinfo{year}{2016}\natexlab{a}.
\newblock \showarticletitle{Using Gaussian Processes for Rumour Stance
  Classification in Social Media}.
\newblock \bibinfo{journal}{\emph{CoRR}}  \bibinfo{volume}{abs/1609.01962}
  (\bibinfo{year}{2016}).
\newblock
\showeprint[arxiv]{1609.01962}
\urldef\tempurl%
\url{http://arxiv.org/abs/1609.01962}
\showURL{%
\tempurl}


\bibitem[\protect\citeauthoryear{Lukasik, Srijith, Vu, Bontcheva, Zubiaga, and
  Cohn}{Lukasik et~al\mbox{.}}{2016b}]%
        {lukasik2016hawkes}
\bibfield{author}{\bibinfo{person}{Michal Lukasik}, \bibinfo{person}{PK
  Srijith}, \bibinfo{person}{Duy Vu}, \bibinfo{person}{Kalina Bontcheva},
  \bibinfo{person}{Arkaitz Zubiaga}, {and} \bibinfo{person}{Trevor Cohn}.}
  \bibinfo{year}{2016}\natexlab{b}.
\newblock \showarticletitle{Hawkes processes for continuous time sequence
  classification: an application to rumour stance classification in twitter}.
  In \bibinfo{booktitle}{\emph{Proceedings of the 54th Annual Meeting of the
  Association for Computational Linguistics (Volume 2: Short Papers)}},
  Vol.~\bibinfo{volume}{2}. \bibinfo{pages}{393--398}.
\newblock


\bibitem[\protect\citeauthoryear{Mei and Eisner}{Mei and Eisner}{2017}]%
        {mei2017neural}
\bibfield{author}{\bibinfo{person}{Hongyuan Mei} {and} \bibinfo{person}{Jason~M
  Eisner}.} \bibinfo{year}{2017}\natexlab{}.
\newblock \showarticletitle{The neural hawkes process: A neurally
  self-modulating multivariate point process}. In
  \bibinfo{booktitle}{\emph{Advances in Neural Information Processing
  Systems}}. \bibinfo{pages}{6754--6764}.
\newblock


\bibitem[\protect\citeauthoryear{Mohler, Short, Brantingham, Schoenberg, and
  Tita}{Mohler et~al\mbox{.}}{2011}]%
        {mohler2011self}
\bibfield{author}{\bibinfo{person}{George~O Mohler}, \bibinfo{person}{Martin~B
  Short}, \bibinfo{person}{P~Jeffrey Brantingham},
  \bibinfo{person}{Frederic~Paik Schoenberg}, {and} \bibinfo{person}{George~E
  Tita}.} \bibinfo{year}{2011}\natexlab{}.
\newblock \showarticletitle{Self-exciting point process modeling of crime}.
\newblock \bibinfo{journal}{\emph{J. Amer. Statist. Assoc.}}
  \bibinfo{volume}{106}, \bibinfo{number}{493} (\bibinfo{year}{2011}),
  \bibinfo{pages}{100--108}.
\newblock


\bibitem[\protect\citeauthoryear{Rizoiu, Lee, Mishra, and Xie}{Rizoiu
  et~al\mbox{.}}{2017}]%
        {rizoiu2017tutorial}
\bibfield{author}{\bibinfo{person}{Marian-Andrei Rizoiu},
  \bibinfo{person}{Young Lee}, \bibinfo{person}{Swapnil Mishra}, {and}
  \bibinfo{person}{Lexing Xie}.} \bibinfo{year}{2017}\natexlab{}.
\newblock \showarticletitle{A tutorial on hawkes processes for events in social
  media}.
\newblock \bibinfo{journal}{\emph{arXiv preprint arXiv:1708.06401}}
  (\bibinfo{year}{2017}).
\newblock


\bibitem[\protect\citeauthoryear{Santosh, Bansal, and Saha}{Santosh
  et~al\mbox{.}}{2019}]%
        {santosh2019can}
\bibfield{author}{\bibinfo{person}{TYSS Santosh}, \bibinfo{person}{Srijan
  Bansal}, {and} \bibinfo{person}{Avirup Saha}.}
  \bibinfo{year}{2019}\natexlab{}.
\newblock \showarticletitle{Can Siamese Networks help in stance detection?}. In
  \bibinfo{booktitle}{\emph{Proceedings of the ACM India Joint International
  Conference on Data Science and Management of Data}}. ACM,
  \bibinfo{pages}{306--309}.
\newblock


\bibitem[\protect\citeauthoryear{Seonwoo, Oh, and Park}{Seonwoo
  et~al\mbox{.}}{2018}]%
        {seonwoo-etal-2018-hierarchical}
\bibfield{author}{\bibinfo{person}{Yeon Seonwoo}, \bibinfo{person}{Alice Oh},
  {and} \bibinfo{person}{Sungjoon Park}.} \bibinfo{year}{2018}\natexlab{}.
\newblock \showarticletitle{Hierarchical {D}irichlet {G}aussian Marked {H}awkes
  Process for Narrative Reconstruction in Continuous Time Domain}. In
  \bibinfo{booktitle}{\emph{Proceedings of the 2018 Conference on Empirical
  Methods in Natural Language Processing}}. \bibinfo{publisher}{Association for
  Computational Linguistics}, \bibinfo{address}{Brussels, Belgium},
  \bibinfo{pages}{3316--3325}.
\newblock
\urldef\tempurl%
\url{https://doi.org/10.18653/v1/D18-1369}
\showDOI{\tempurl}


\bibitem[\protect\citeauthoryear{Sha, Hasan, Mohler, and Brantingham}{Sha
  et~al\mbox{.}}{2020}]%
        {sha2020dynamic}
\bibfield{author}{\bibinfo{person}{Hao Sha}, \bibinfo{person}{Mohammad~Al
  Hasan}, \bibinfo{person}{George Mohler}, {and} \bibinfo{person}{P~Jeffrey
  Brantingham}.} \bibinfo{year}{2020}\natexlab{}.
\newblock \showarticletitle{Dynamic topic modeling of the COVID-19 Twitter
  narrative among US governors and cabinet executives}.
\newblock \bibinfo{journal}{\emph{arXiv preprint arXiv:2004.11692}}
  (\bibinfo{year}{2020}).
\newblock


\bibitem[\protect\citeauthoryear{Veyseh, Ebrahimi, Dou, and Lowd}{Veyseh
  et~al\mbox{.}}{2017}]%
        {veyseh2017temporal}
\bibfield{author}{\bibinfo{person}{Amir Pouran~Ben Veyseh},
  \bibinfo{person}{Javid Ebrahimi}, \bibinfo{person}{Dejing Dou}, {and}
  \bibinfo{person}{Daniel Lowd}.} \bibinfo{year}{2017}\natexlab{}.
\newblock \showarticletitle{A temporal attentional model for rumor stance
  classification}. In \bibinfo{booktitle}{\emph{Proceedings of the 2017 ACM on
  Conference on Information and Knowledge Management}}. ACM,
  \bibinfo{pages}{2335--2338}.
\newblock


\bibitem[\protect\citeauthoryear{Xiao, Yan, Yang, Zha, and Chu}{Xiao
  et~al\mbox{.}}{2017}]%
        {xiao2017modeling}
\bibfield{author}{\bibinfo{person}{Shuai Xiao}, \bibinfo{person}{Junchi Yan},
  \bibinfo{person}{Xiaokang Yang}, \bibinfo{person}{Hongyuan Zha}, {and}
  \bibinfo{person}{Stephen~M Chu}.} \bibinfo{year}{2017}\natexlab{}.
\newblock \showarticletitle{Modeling the intensity function of point process
  via recurrent neural networks}. In \bibinfo{booktitle}{\emph{Thirty-First
  AAAI Conference on Artificial Intelligence}}.
\newblock


\bibitem[\protect\citeauthoryear{Zubiaga, Kochkina, Liakata, Procter, and
  Lukasik}{Zubiaga et~al\mbox{.}}{2016a}]%
        {zubiaga2016stance}
\bibfield{author}{\bibinfo{person}{Arkaitz Zubiaga}, \bibinfo{person}{Elena
  Kochkina}, \bibinfo{person}{Maria Liakata}, \bibinfo{person}{Rob Procter},
  {and} \bibinfo{person}{Michal Lukasik}.} \bibinfo{year}{2016}\natexlab{a}.
\newblock \showarticletitle{Stance classification in rumours as a sequential
  task exploiting the tree structure of social media conversations}.
\newblock \bibinfo{journal}{\emph{arXiv preprint arXiv:1609.09028}}
  (\bibinfo{year}{2016}).
\newblock


\bibitem[\protect\citeauthoryear{Zubiaga, Kochkina, Liakata, Procter, Lukasik,
  Bontcheva, Cohn, and Augenstein}{Zubiaga et~al\mbox{.}}{2018}]%
        {zubiaga2018discourse}
\bibfield{author}{\bibinfo{person}{Arkaitz Zubiaga}, \bibinfo{person}{Elena
  Kochkina}, \bibinfo{person}{Maria Liakata}, \bibinfo{person}{Rob Procter},
  \bibinfo{person}{Michal Lukasik}, \bibinfo{person}{Kalina Bontcheva},
  \bibinfo{person}{Trevor Cohn}, {and} \bibinfo{person}{Isabelle Augenstein}.}
  \bibinfo{year}{2018}\natexlab{}.
\newblock \showarticletitle{Discourse-aware rumour stance classification in
  social media using sequential classifiers}.
\newblock \bibinfo{journal}{\emph{Information Processing \& Management}}
  \bibinfo{volume}{54}, \bibinfo{number}{2} (\bibinfo{year}{2018}),
  \bibinfo{pages}{273--290}.
\newblock


\bibitem[\protect\citeauthoryear{Zubiaga, Liakata, Procter, Hoi, and
  Tolmie}{Zubiaga et~al\mbox{.}}{2016b}]%
        {zubiaga2016analysing}
\bibfield{author}{\bibinfo{person}{Arkaitz Zubiaga}, \bibinfo{person}{Maria
  Liakata}, \bibinfo{person}{Rob Procter}, \bibinfo{person}{Geraldine Wong~Sak
  Hoi}, {and} \bibinfo{person}{Peter Tolmie}.}
  \bibinfo{year}{2016}\natexlab{b}.
\newblock \showarticletitle{Analysing how people orient to and spread rumours
  in social media by looking at conversational threads}.
\newblock \bibinfo{journal}{\emph{PloS one}} \bibinfo{volume}{11},
  \bibinfo{number}{3} (\bibinfo{year}{2016}), \bibinfo{pages}{e0150989}.
\newblock


\end{thebibliography}
\clearpage

\end{document}